\documentclass[iop,jcp,twocolumn,superscriptaddress]{revtex4-1}

\usepackage{slantsc}
\usepackage{lmodern}
\usepackage{graphicx}
\usepackage{dcolumn}
\usepackage{amsmath}
\usepackage{bm}        
\usepackage{amssymb}   
\usepackage{xcolor}

\begin{document}


\title{A semiclassical nonequilibrium Green's Function approach to electron transport in systems exhibiting electron-phonon couplings }


\author{Maicol A. Ochoa}
\email[]{maicol@umd.edu}
\affiliation{Department of Chemistry and Biochemistry, University of Maryland, College Park, Maryland 20742, USA}
\affiliation{Institute for Physical Science \& Technology, University of Maryland, College Park, Maryland 20742, USA}


\date{\today}

\begin{abstract}
  We formulate a semiclassical theory for electron transport in open quantum systems with electron-phonon interactions adequate for situations when the system's phonon dynamics is comparable with the electron transport timescale. Starting from the Keldysh non-equilibrium Green's function formalism we obtain equations of motion for the retarded and lesser electronic Green's functions including contributions due to the phonon dynamics up to second order in the electron-phonon coupling strength. The resulting equations assume that the system's phonon follow classical time-local dynamics with delta-correlated noise. We apply our method to the study of the charging/discharging of a periodically driven quantum dot, and a three-level model for a single-electron pump, analyzing the signatures in the transient current, electron population and process performance of the phonon dynamics. For these systems, \textcolor{black}{we adopt the fluctuation-dissipation theorem} and consider external harmonic driving of the phonon at frequencies comparable with the electron modulation, and different scenarios, varying electron-phonon coupling strength, coupling to the electron part of the system, and in phase and anti-phase driving. Our results illustrate that our method provides an efficient protocol to describe the effects of nuclear motion in ultrafast transient phenomena.
\end{abstract}


\maketitle

\section{Introduction}

Understanding electron transport in nanoscale devices is fundamental for developing novel technologies and their operational control. In this regime, quantum confinement\cite{ochoa2024single}, environmental noise, and many-body interactions\cite{ochoa2025atomistic} such as electron-phonon couplings can significantly alter the electron dynamics. The latter is typical in nanowires\cite{jin2006three,zhang2010atomistic}, nanotubes\cite{perebeinos2005electron,roche2007charge,adinehloo2023phonon},  quantum dots\cite{ueda2007nonequilibrium,maier2011charge,krzywda2021interplay,mathe2022quantum}, semiconductors\cite{zhou2020direct,ochoa2020extracting,zhou2022anomalous,ochoa2023gasb}, 2D materials\cite{sangwan2018electronic,kaasbjerg2013acoustic,kaasbjerg2012phonon}, topological insulators\cite{vannucci2020conductance,akhoundi2023impact}, and molecular junctions\cite{galperin2007molecular,franco2007robust,hihath2012electron,troisi2006molecular,hartle2009vibrational,hartle2011vibrational,ballmann2012experimental,evers2020advances,craven2016electron,craven2017electron,craven2017electrothermal,craven2023electron,cui2021heat,hu2018lessons,mejia2018signatures,venkataraman2011semiclassical,bi2020electron,kosov2017non,kosov2017waiting}, facilitating applications such as thermoelectricity\cite{jiang2015phonon,esposito2015efficiency,xi2022perspective}, nanoelectromechanical sensing\cite{ferrari2023nanoelectromechanical,ochoa2019generalized,ochoa2020optimal}, spectroscopy\cite{galperin2004inelastic,galperin2009raman}, flexoelectronics\cite{lou2022flexoelectronic}, and plasmonics\cite{mondal2022coupling,mondal2022strong}. Several methods for theoretically modeling these systems, such as Boltzmann transport equations\cite{zhang2010atomistic} and Nonequilibrium Green's Functions (NEGFs)\cite{jauho1994time}, are frequently used. The former describes electrons and phonons classically, while the latter treats them as quantum particles. In comparison, NEGFs are more accurate but are computationally more expensive to solve. As a result, there is a need to develop efficient and accurate quantum-classical methods for these systems.

In nanoscale and molecular junctions, accurate modeling of electron-phonon couplings is essential to understanding inelastic transport\cite{segal2002conduction,troisi2003vibronic,galperin2004inelastic, simine2014electron, agarwalla2016reconciling, galperin2005molecular,galperin2006resonant,galperin2007molecular}, negative differential conductivity\cite{hartle2011resonant}, and energy dissipation\cite{quan2021impact,ochoa2018quantum,ganguly2024transport,ochoa2022quantum,ochoa2024lossy}. Studies in nonadiabatic phenomena\cite{kershaw2020non}, current-induced chemical reactions, and electronic friction\cite{dou2017born,miao2017vibrational,dou2018perspective,bi2024electronic,chen2018current,chen2019electronic} resort to timescale separation of the nuclear and electron dynamics, following the Born-Oppenheimer approximation, leading to a transport picture where the nuclei are slow classical particles moving in the electronic potential. Methods of this type include master equations\cite{dou2018universal,wilner2014nonequilibrium}, path integral methods\cite{thornber1971linear,muhlbacher2008real,simine2013path}, time-dependent Hartree-Fock theory\cite{wang2009numerically}, time-dependent density functional theory\cite{chen2018time,bostrom2019electron}, renormalization group methods\cite{tsai2006renormalization,girotto2023dynamical,yang2023time}, and NEGF with the Ehrenfest approximation\cite{esposito2015quantum,preston2020current,kershaw2020non,preston2022emergence}, or in the Hubbard-operator description\cite{chen2018current,chen2019electronic}.

In this paper, we introduce a semiclassical method for electron transport in quantum systems with electron-phonon couplings and fast nuclear or phonon dynamics based on the NEGF  technique, modified to allow for classical or phenomenological equations of motion for the system's phonons. We derive closed differential equations for the retarded and lesser projections of the electronic Green function incorporating the electron-phonon couplings up to second order in the coupling strength. The resulting equations depend on the first and second moments of the phonon position operators, which we chose to estimate via stochastic classical equations, resulting in semiclassical self-energies. \textcolor{black}{Compared with other methods, such as those employing the Ehrenfest approximation\cite{haug2008quantum}, which involves separating fast quantum and slow macroscopic variations, our method is valid for arbitrary regimes in classical phonon dynamics. A common realization of the Ehrenfest approximation involves introducing Wigner coordinates and performing a truncated gradient expansion. When the phonons correspond to the nuclear motions of the system, the Ehrenfest approximation is effectively implemented by considering that the atomic dynamics is slow compared to the electron tunneling\cite{preston2022emergence}. This timescale separation is not needed if the dynamics of the phonon position and momentum expectation values are obtained directly from the Ehrenfest theorem, without the Ehrenfest approximation. In this case, the Ehrenfest dynamics requires the exact evaluation of the quantum dynamics to recover the first moments in position and momentum. Methods of this type include the Ehrenfest-Keldysh approach\cite{hopjan2018molecular,preston2022emergence}. Our proposed semiclassical method relies on the first and second position moments and remains valid even in cases where the nuclear dynamics are comparable to the electron transport rates.} However, it is perturbative in the electron-phonon coupling strength. We investigate the dynamics of a driven quantum dot and a model for a single-electron pump coupled to harmonically driven phonons at rates comparable with the electron dynamics. In both cases, phonon couplings result in additional smearing of the electric currents and sensitive variations in the electron densities in the system. 

The organization of the paper is as follows. In Sec. \ref{sec:theory}, we introduce the system's model and derive the generic form of the semiclassical equations of motion for the retarded and lesser electronic Green's Functions functionally dependent on the phonon dynamics. Then, in Sec. \ref{sec:numerics}, we apply this formalism and investigate the impact on the electron current and electron population in a periodically driven quantum dot and a model for a single-electron pump, assuming different couplings and phonon driving regimes. We conclude in Sec. \ref{sec:conclusion}

\section{Semiclassical theory}\label{sec:theory}

In this section we derive the semiclassical dynamical equations describing electron transport through a system with electron-phonon couplings.  In the following we work in atomic units, such that $\hbar = 1$.

\subsection{Hamiltonian}

Our system of interests is a generic open quantum systems with electron and phonon degrees of freedom defined by a tripartite Hamiltonian $\hat H$ 
\begin{align}
  \hat H = \hat H_{\rm S} + \hat H_{\rm B} + \hat H_{\rm SB},\label{eq:GenHam}
\end{align}
where $\hat H_{\rm S}$,  $\hat H_{\rm B}$ and $\hat H_{\rm SB}$  are correspondingly the Hamiltonians for the system, the baths representing the environment, and  the system-bath coupling terms. The generic system Hamiltonian $\hat H_{\rm S}$ in second quantized form reads
\begin{align}
  \hat H_{\rm S} ={}& \sum_i \varepsilon_i \hat d_i^\dagger \hat d_i + \sum_{i \neq j} t_{ij} \hat d_i^\dagger \hat d_j\notag \\
                    &+ \sum_{i,\alpha} M_i^\alpha \hat d_i^\dagger \hat d_i (\hat a_\alpha^\dagger+\hat a_\alpha) + \hat H_{\rm ph},\label{eq:Hsys}
\end{align}
with $\hat d_i^\dagger$ ($\hat d_i$) the creation (annihilation) operator for an electron with energy $\varepsilon_i$, $t_{ij}$ is the coupling energy for an electron tunneling from the $j$-th to the $i$-th energy level, $\hat a_\alpha^\dagger$ ($\hat a_\alpha $) is the mode creation (annihilation) operator for the $\alpha$-th phonon, $M_i^\alpha$ is the coupling strength between the $i$-th level and the $\alpha$-th phonon.  \textcolor{black}{In Eq.\ \eqref{eq:Hsys}, we include the Hamiltonian $\hat H_{\rm ph}$ to account for the primary phonons.}

 The surroundings consist of free-electron and phononic reservoirs, such that   
\begin{align}
  \hat H_{\rm B} ={}& \sum_K \sum_{j \in K} \varepsilon_j^{(K)} \hat c_{j,K}^\dagger \hat c_{j,K} + \hat H_{\rm P},
\end{align}
where $\hat c_j^\dagger$ ($\hat c_j$) is the creation (annihilation) operator for an electron in the $K$-th electronic bath with energy $\varepsilon_j^{(K)}$, and $\hat H_{\rm P}$ is the unspecified Hamiltonian for the phonon bath. Moreover, the coupling between the system and the baths is given by
\begin{align}
  \hat H_{\rm SB} = {}& \sum_K \sum_i \sum_{j \in K} W_{ij}^{(K)} \hat d_i^\dagger \hat c_{j,K} +W_{ij}^{(K) *} \hat c_{j,K}^\dagger  \hat d_i + \hat H_{\rm ph-P}, 
\end{align}
with coupling energies $W_{ij}^{(K)}$ between the $i$-th level in the system and the $j$-th level in the $K$-th bath. $\hat H_{\rm ph-P}$ is the Hamiltonian collectively representing the primary phonon couplings with the environment. In the above expressions, we deliberately refrained from defining $\hat H _{\rm ph}$, $\hat H_{\rm P}$, and $\hat H_{\rm ph-P}$, as we will treat their contributions to the system dynamics classically.

\subsection{Electron transport and correlations}

 Within the nonequilibrium Green's Function formalism (NEGF), the collection of  two-time Green Functions (GF) between the $i$th and $j$th electrons 
\textcolor{black}{
\begin{align}
  G_{ij}(\tau,\tau') ={}& -i \langle \hat d_{i,\mathcal{I}}(\tau) \hat d_{j,\mathcal{I}}^\dagger (\tau') \rangle_c,\label{eq:GF}\\
  ={}& -i \frac{{\rm Tr} \left\{ \rho_{o} T_c \hat d_{i,\mathcal{I}}(\tau) \hat d_{j,\mathcal{I}}^\dagger (\tau') \hat S_C \right\}}{{\rm Tr}\left\{\rho_{o}T_c \hat S_C\right\}},\label{eq:GF2}
\end{align}
}carry the information needed to obtain the electron dynamics, coherence and population. \textcolor{black}{In Eqs.\ (5) and (6), $\tau$ and $\tau'$ are Keldysh contour variables, with the contour defined as the smooth path $C$ starting and ending at $\tau=-\infty$, reaching $\tau$ and $\tau'$; $T_c$ is the contour ordering operator, and the subindex $\mathcal{I}$ indicates an operator in the interaction picture representation. The density matrix $\rho_o$ is the product $\rho_L^{eq}\otimes\rho_R^{eq}\otimes\rho_{\rm S}^o$, consisting of the uncoupled density matrices for the reservoirs and the system at equilibrium. The latter also factors as the product of the equilibrium density matrices for the electron and phonon system parts  $\rho_{\rm S}^o =\rho_{\rm elec} \otimes \rho_{\rm phon}$. The S-matrix $\hat S_C$ is defined by
\begin{align}
  \hat S_C ={}& T_c \exp \left( i \int_C \hat V_{\mathcal{I}}(\tau)d\tau \right),
\end{align}
in terms of the interacting Hamiltonian  $\hat V$. The S-matrix introduces adiabatically system-bath and electron-phonon interactions, conditional to having all time-dependent processes starting at $t\ge 0$ \cite{haug2008quantum}.} For the system defined in Eq.\ \eqref{eq:GenHam} we consider
\begin{align} 
  \hat V ={}&\sum_{i,\alpha} M_i^\alpha \hat d_i^\dagger \hat d_i (\hat a_\alpha^\dagger+\hat a_\alpha)+\hat H_{\rm SB}.
\end{align}

In the following, we will omit the index $\mathcal{I}$ to simplify notation, understanding that all the operators in the GFs are in the interaction picture unless otherwise stated.

The transient current between the system and the $K$-th bath follows from the retarded and lesser projections of the full electronic Green Function $G$ according to the expression\cite{jauho1994time,ochoa2015pump} 
\begin{align}
  I_{K}(t) = - {\rm Im} &\Big( {\rm Tr} \Big\{G^<(t,t)\Gamma_K  \notag \\
  &\left. \left. + \int \frac{d \varepsilon}{2 \pi} G^r(t,\varepsilon)\Gamma_{K} f(\varepsilon, \mu_K)\right\} \right),\label{eq:It}
\end{align}
where the trace ${\rm Tr}$ is taken on the electronic degrees of freedom, $f(\varepsilon, \mu_K)$ is the Fermi function, $\mu_K$ is the chemical potential of the $K$-th reservoir, and $[\Gamma_K]_{ij} = 2 \pi \sum_{l \in K} W_{il}^{(K)*} W_{jl}^{(K)} \delta(\varepsilon_l^{(K)}- \varepsilon $), \textcolor{black}{ and  $G^r(t,\varepsilon)$ is the $G^{r}$ Fourier transform
\begin{equation}
  \label{eq:Grdef}
  G^{r}(t,\varepsilon) = \int d t' G^r(t, t') e^{-i \varepsilon (t' - t)}.
\end{equation}
}

For the model defined in Eq.\ \eqref{eq:GenHam}, the Heisenberg equation of motion for $\hat d_i$ is given by
\begin{align}
  i \frac{d}{dt}\hat d_i  ={}& \varepsilon_i \hat d_i + \sum_{j \neq i } t_{ij} \hat d_j \notag\\
  &+ \sum_\alpha M_i^\alpha \hat d_i \hat X_\alpha + \sum_{K, j \in K} W_{ij}^{(K)}\hat c_{j,K}\; ,   
\end{align}
with  $\hat X_\alpha = \hat a^\dagger_\alpha+\hat a _\alpha $; and the consequently equation of motion in the Keldysh contour for $G_{ij}(\tau,\tau')$, Eq.\ \eqref{eq:GF}, is
\begin{align}
  i \frac{d}{d\tau}G_{ij}(\tau,\tau') ={}&\delta(\tau, \tau')\delta_{ij}+ \varepsilon_i G_{ij}(\tau,\tau') + \sum_{l \neq i } t_{il}  G_{lj}(\tau,\tau') \notag\\
                                         &+ \sum_\alpha M_i^\alpha \left (-i \langle \hat X_\alpha(\tau) \hat d_i(\tau) \hat d_j^\dagger(\tau') \rangle_c \right )\notag\\
  & +\sum_{K, l \in K } W_{il}^{(K)}\left(-i \langle \hat c_{l,K}(\tau) \hat d^\dagger_j(\tau) \rangle_c\right)  .\label{eq:dtGF}
\end{align}

In the absence of phonon-electron coupling (i.e.  $M_i^\alpha = 0$), Eq.\ \eqref{eq:GF} coincides with the known form for the GF equation of motion for a set of noninteracting electron levels.  Invoking the Langreth rules, one can project Eq.\ \eqref{eq:dtGF} to obtain the evolution equation for $G^r$. This also requires a special consideration of the correlation term proportional to the phonon-electron coupling energy.
 Introducing the notation
\begin{align}
  F^\alpha_{ij}(\tau,\tau')=& -i \langle \hat X_\alpha (\tau)  \hat d_i(\tau) \hat d_j^\dagger(\tau') \rangle_c,\label{eq:Fij}
\end{align}
and expanding up the S-matrix $\hat S_C$, 
\begin{align}
\textcolor{black}{
\hat S_C^{(2)} = }&\textcolor{black}{ I + i \int_C d\tau \hat V(\tau) - \frac{1}{2} T_c \int_C d\tau_1 \int_C d\tau_2  \hat V(\tau_1) \hat V(\tau_2),}
\intertext{we evaluate the correlation in Eq.\ \eqref{eq:Fij} and find}
             i F^{\alpha}_{ij}&(\tau,\tau')=\langle \hat X_\mathcal{\alpha}(\tau) \rangle_c \langle \hat d_i(\tau)\hat d^\dagger (\tau') \rangle_c\notag\\
  +& i \sum_{l,\beta} M_l^\beta \int d \tau_1 \langle \hat X_\alpha (\tau) \hat X_\beta (\tau_1)\rangle_c \langle \hat d_i (\tau) \hat d_j^\dagger(\tau') \hat n_l(\tau_1)\rangle_c\label{eq:Faij}
\end{align}
where we introduced the electron number operator $\hat n_l(\tau_1) = \hat d_l^\dagger(\tau_1) \hat d_l(\tau_1) $. The integrand in the second term in Eq.\ \eqref{eq:Faij} can be further simplified utilizing Wick's theorem after \textcolor{black}{noting that the density matrix $\rho_o$ in Eq.\ \eqref{eq:GF2} is quadratic and uncorrelated}
\begin{align}
  \langle \hat d_i (\tau) \hat d_j^\dagger(\tau') \hat n_l(\tau_1)\rangle_c =& i G_{ij}(\tau,\tau')\langle\hat n_l(\tau_1) \rangle_c\notag \\
  &\hspace{0.5cm}+ G_{il}(\tau, \tau_1)G_{lj}(\tau_1, \tau').
\end{align}

The above results and considerations allow us to write the equation of motion for $G$ in a concise form. For a system consisting of $m$ phonons and $n$ electron levels, we define the $(n,m)$ matrix $M$
\begin{equation}
  \label{eq:Mmat}
  [M]_{i,\alpha} = M_i^\alpha,
\end{equation}
such that its row vector 
\begin{align}
  M_i =& (M_i^{\alpha_1},\dots, M_i^{\alpha_m})
\end{align}
lists the electron-phonon coupling terms between the $i$th electron level and the primary phonon manifold. We also introduce the $(m, m)$ matrix  $K$
\begin{equation}
  \label{eq:Kmat}
  [K]_{\alpha,\beta}(\tau,\tau_1) = \langle \hat X_\alpha (\tau) \hat X_\beta(\tau_1) \rangle_c ,
\end{equation}
consisting of the primary-phonon position correlations, and the vectors
\begin{align}
  \vec{\rm x}(\tau) =& ( \langle \hat X_{\alpha_1} (\tau) \rangle_c,\dots,\langle \hat X_{\alpha_m} (\tau) \rangle_c ),\\
  \vec{\rm n}(\tau)=& ( \langle  \hat n_{1} (\tau) \rangle_c,\dots,\langle \hat n_{n} (\tau) \rangle_c ).
\end{align}
Thus, the equation of motion in Eq.\ \eqref{eq:dtGF} for the matrix of electronic Green Functions $G$ reads\footnote{\textcolor{black}{ We note that the term ${\rm diag}\left[M\dot \vec x (\tau_1)\right]$ in Eq.\ \eqref{eq:dtGmat} represents the Hartree term, which can be included in the self-energy $\Sigma_{\rm ph}$ by multiplying this term by  $\delta(\tau,\tau_1) $. Both forms are equivalent.}}
\begin{align}
  \label{eq:dtGmat}
  i \frac{d}{d\tau} G(\tau,\tau') &= \delta(\tau,\tau') I + \left(\textcolor{black}{\hat H_e}+{\rm diag}[M \cdot \vec{\rm x}(\tau)]\right) G(\tau, \tau')+\notag\\
                               &+i \int_c d \tau_1\, {\rm diag}\left( M K(\tau,\tau_1)M^T \cdot \vec{\rm n}(\tau_1) \right) G(\tau,\tau')\notag\\
                               &+ \int_c d \tau_1 \Big(\Sigma_{\rm ph}(\tau,\tau_1)+\Sigma_{\rm e}(\tau,\tau_1)\Big) G(\tau_1,\tau'),
\end{align}
\textcolor{black}{where $\hat H_e = \sum_i \varepsilon_i \hat d_i^\dagger \hat d_i + \sum_{i\ne j}t_{ij}\hat d_i^\dagger \hat d_j$ and} the electron-electron and phonon-electron self-energies are correspondingly given by
\begin{align}
  \left[ \Sigma_{\rm e}(\tau, \tau') \right]_{i,j} & = \sum_K \sum_{k} W^{(K)}_{ik} W^{(K)\, *}_{jk}   g_{ij}^{(K)}(\tau, \tau'), \label{eq:selfE} \\
  \Sigma_{\rm ph}(\tau, \tau') & = \left(M K(\tau, \tau')M^T\right) \odot G(\tau,\tau'), \label{eq:selfPH}
\end{align}
with $g_{ij}^{(K)}(\tau, \tau') = - i \langle \hat c^\dagger_{i,K}(\tau) \hat c_{j,K}(\tau') \rangle_c  $  and $\odot$ denoting the Hadamard matrix product. Since the Hadamard product is commutative, the self-energy $\Sigma_{\rm ph}$ is well-defined irrespective of the contour variable considered in the evaluation of the equation of motion ( See Ref. for further discussion \cite{ochoa2014non}). We also note that $ \Sigma_{\rm ph}$ depends explicitly on the electron Green's Function implying that Eq.\ \eqref{eq:dtGmat} is nonlinear in $G$. Moreover, to determine the correlation functions in $K$, one formulates analog equation of motion for the phonon degrees of freedom which include self-energy terms proportional to the electron GF\cite{galperin2004inelastic}, adding to the dynamical complexity of this system. As a computationally efficient alternative, we formulate a quantum-classical approach to the evolution of the $G$ matrix in the next section.

\subsection{ Semiclassical lesser and retarded electron Green's Functions}

We adopt a mixed quantum-classical description of the transport problem considering classical or phenomenological equations of motion for the primary phonon degrees of freedom. Specifically, we assume that a mapping for each observable $\hat X_\alpha$ into a real variable $x_\alpha$ exists, and the $x_\alpha$ dynamics follows a stochastic differential equation.  In particular, we restrict our analysis to variables with time-local dynamics and delta-correlated noise. Thus, in the semiclassical representation of $\vec{\rm x}$ and $K$ in Eq. \eqref{eq:dtGmat} we substitute the expectation values  such that
\begin{align}
  \langle X_\alpha(\tau)\rangle \equiv& \overline{ x_\alpha (t)}\\
  \langle \hat X_\alpha (\tau) \hat X_\beta (\tau') \rangle \equiv& \delta(t,t')\overline{ x_\alpha(t)x_\beta(t')} ,
\end{align}
where the overline represents the classical or phenomenological mean value.  \textcolor{black}{Delta-correlated noise is an effective approximation in the systems for which the noise source appears white  over the energy range relevant to the system. This is the case with many nanomechanical resonators, trapped atoms in optical cavities, molecular vibrations and rotations in solution, and microcantilevers in contact with a thermal bath, among others.} With this assumption, the semiclassical version of the self-energy $\tilde \Sigma_{\rm ph}$ in Eq.\ \eqref{eq:selfPH}, consists of a classical term, the product $M K(t,t) M^T \equiv M K(t) M^T$ and a quantum electronic contribution $G(\tau, \tau')$. As an additional consideration, we will assume the wide band approximation (WBA), such that the elements in $\Gamma_K$ are energy independent.

These assumptions on the phonon dynamics allows us to compute the equation of motion for the retarded projections of the electron Green's Function $G^r$ 
\begin{align}
  i\frac{d}{dt}G^r&(t,\varepsilon) = I +\left[\textcolor{black}{\hat H_e(t)} - \varepsilon I + {\rm diag }[M \cdot \vec{\rm x}(t)]\right] G^r(t, \varepsilon)\notag\\
  &\hspace{1.4cm}+i{\rm diag}[M K(t) M^T \cdot \vec{\rm n}(t)] G^r(t, \varepsilon)\notag\\
                  &\hspace{2cm} +\tilde \Sigma_{\rm ph}^{r}(t) G^r(t,\varepsilon) - i \frac{\Gamma}{2} G^r(t,E), \label{eq:dGr}       
\end{align}
where $\Gamma = \sum_K \Gamma_K$.

We derive the equation of motion for the semiclassical lesser Green's Function at equal times, $G^<(t,t)$, considering first the Keldysh form for $G^<$
\begin{align}
  G^<(t,t)=\int dt_1\int dt_2G^r(t, t_1)&\Big(\Sigma_{\rm e}^<(t_1, t_2)\notag\\
  &\left.+ \Sigma_{\rm ph}^<(t_1, t_2)\right)G^a(t_2, t),
\end{align}
differentiating and using Eq.\ \eqref{eq:dGr} in the standard form. The result reads
\begin{align}
  i \frac{d}{dt}G^<(t,t)=& i \int \frac{d\varepsilon}{2 \pi} \sum_K f_K(\varepsilon)\Big( \Gamma_K G^a(\varepsilon, t)- G^r(t, \varepsilon) \Gamma_K \Big) \notag \\
                         & + \tilde \Sigma^<_{\rm ph}(t) G^a(\varepsilon,t)-G^r(t,\varepsilon) \tilde \Sigma^<_{\rm ph}(t) \notag\\
                         &+[\hat H_e(t)+ {\rm diag}[M\cdot \vec{\rm x}(t)], G^<(t,t)] \notag\\
                         &+\tilde \Sigma^r_{\rm ph}(t)G^<(t,t)-G^<(t,t)\tilde \Sigma^a_{\rm ph }(t) \notag\\
  &+i \left\{ {\rm diag}[MK(t)M^T \cdot \vec{\rm n}(t)] - \frac{1}{2}\sum_K \Gamma_K, G^<(t,t) \right\},  \label{eq:dGless}
\end{align}

where $\{A,B\} = AB +BA $.  Equations \eqref{eq:dGr} and \eqref{eq:dGless}, and the dynamical equations for the phonon degrees of freedom, constitute the semiclassical model from which one can calculate the transient current in Eq.\ \eqref{eq:It}, and the system's electronic density matrix.

\section{Numerical models}\label{sec:numerics}

In this section we apply the theory in Sec. \ref{sec:theory} to study electron transport in systems with phonon couplings under different drivings. In particular, we consider the effect of electron-phonon couplings in a periodically driven quantum dot, and a model for a single-electron pump.  \textcolor{black}{We assume that primary phonons are classical one-dimensional harmonic oscillators with friction, in contact with a thermal bath, and ocassionally driven by a periodic external field.} Specifically, each phonon coordinate $x_\alpha$ may follow a Langevin equation\cite{nitzan2024chemical,yaghoubi2017energetics}
\begin{align}
  m_\alpha \frac{d^2 x_\alpha}{dt^2}+ \gamma \frac{dx_\alpha}{dt} + m_\alpha \omega_\alpha x_\alpha& = \notag\\
  A_o &\cos(\omega_d t)+ \sqrt{2 \gamma_\alpha k_B T_\alpha} \zeta(t), \label{eq:mdtx}
\end{align}
\begin{figure}[th]
  \centering
  \includegraphics[scale=0.6]{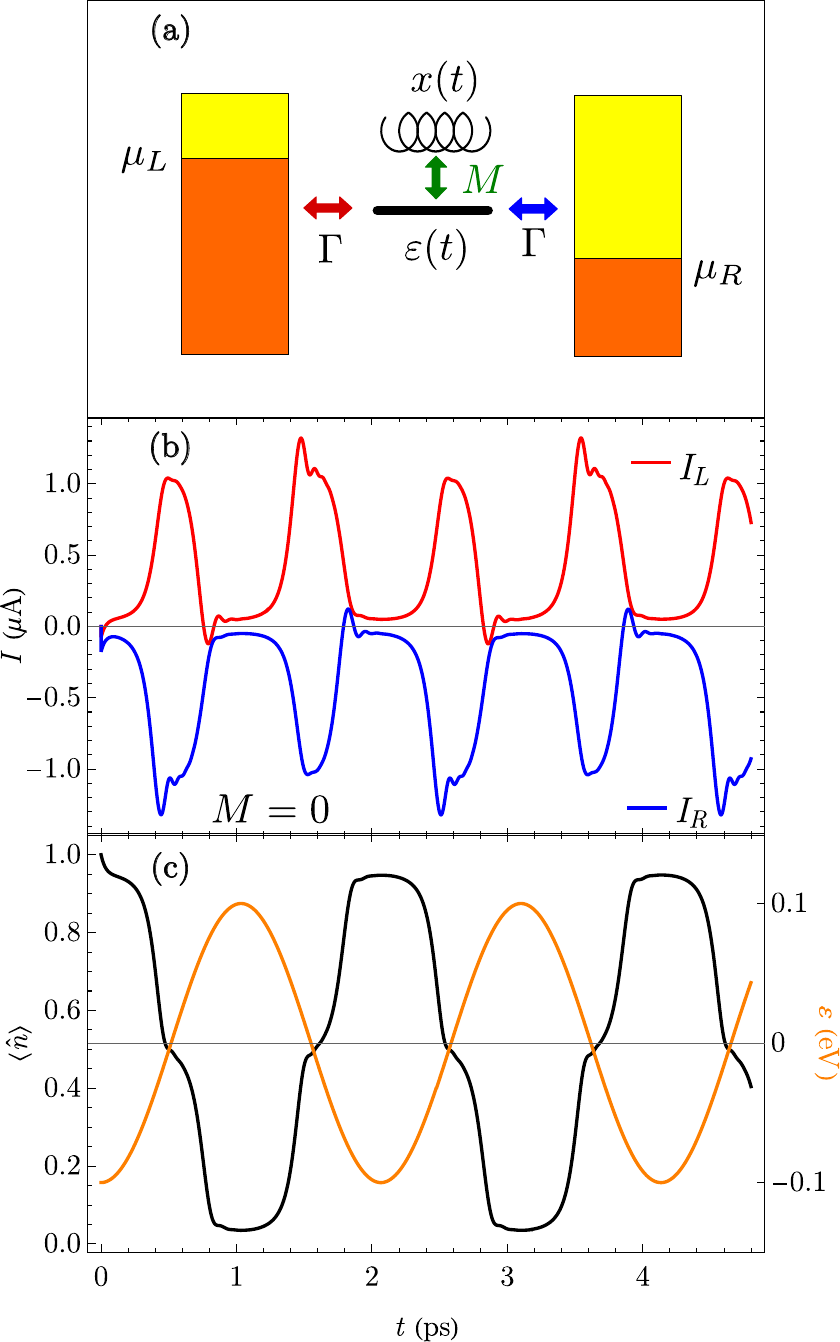}
  \caption{Periodically-driven single dot. (a) Schematic representation of the single dot, with time-dependent energy $\varepsilon(t)$, interacting with two fermionic reservoirs, coupled to a phonon with strength $M$. The left reservoir chemical potential $\mu_L$ is assumed higher than the chemical potential on the right reservoir $\mu_R$, and both are kept at the same temperature. (b) Left ($I_L$, red) and right ($I_R$, blue) current as function of time (c) electron level population(black, left axis) versus level energy (orange, right axis). In this model the phonon is decoupled from the level ($M=0$), $\mu_R =0.5$ eV, $\mu_L = -0.5$ eV, $T = 10$ K, $\Gamma_L = \Gamma_R = \Gamma = 0.01$ eV, $\varepsilon_o =-1.0$ eV, $\omega_d = 2 $ meV.}
  \label{fig:dot}
\end{figure}

\begin{figure}[th]
  \centering
  \includegraphics[scale=0.6]{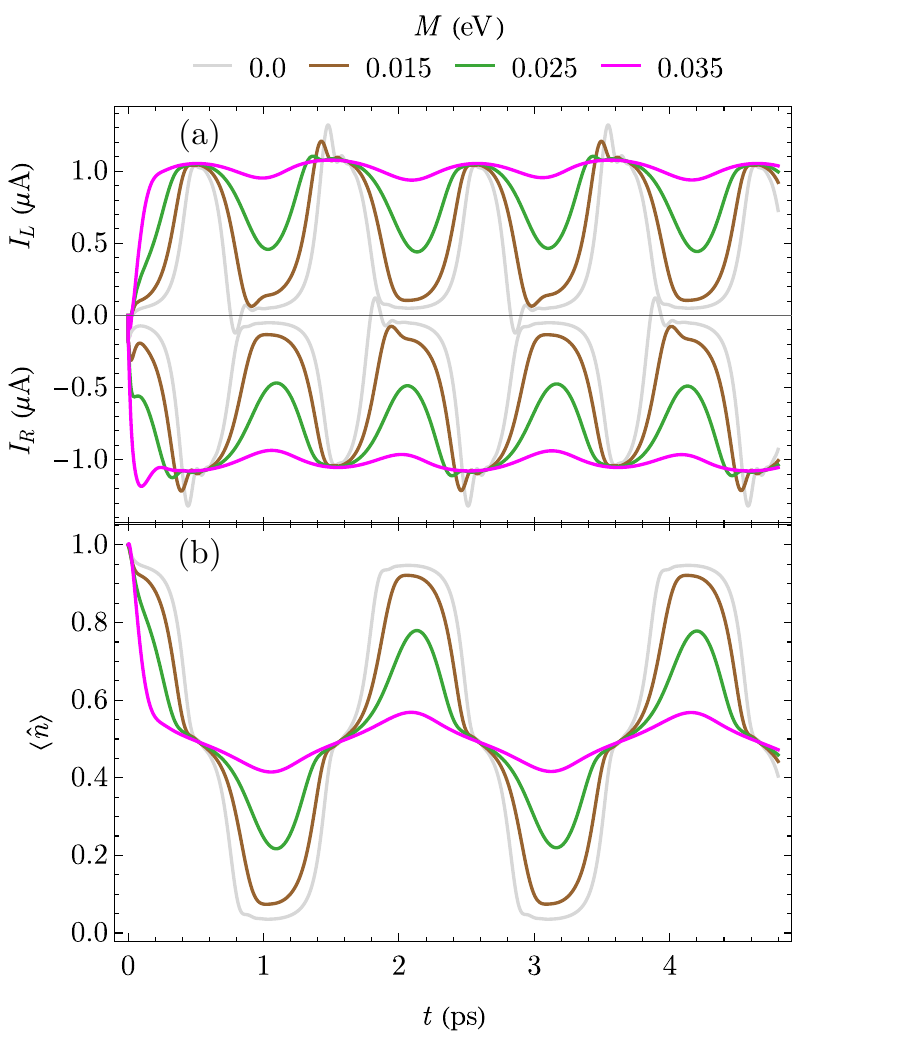}
  \caption{Periodically-driven single dot coupled to a driven phonon, as sketched in Fig. \ref{fig:dot}(a). (a) Left ($I_L$) and right ($I_R$) currents as function of time for various coupling strengths $M$  (b) electron level population as a function of time. In Figs.\ (a) and (b) $M=0.0$ eV (light gray), $M = 0.015$ eV (brown), $M = 0.025$ eV (light blue), $M = 0.035$ eV (purple). \textcolor{black}{The system's phonon is driven by periodically with frequency $\omega_d = 2 $ meV and $\phi = 0$, with $c_1 = 1.89$ and $c_2 =35.71$}. Other parameters are as in Fig.\ \ref{fig:dot} }
  \label{fig:dotM}
\end{figure}

\noindent where $m_\alpha$ is the phonon mass, $\omega_\alpha$ is the phonon characteristic frequency,  $\gamma$ is the friction coefficient, $k_B$ is the Boltzmann constant, and $T_\alpha$ is the local temperature. Equation \eqref{eq:mdtx} includes the possibility of a periodic external driving of the phonon with a frequency $\omega_d$ and amplitude of modulation $A_o$. Moreover, we model thermal fluctuations by a delta correlated random variable $\zeta(t)$ of zero mean (i.e. $\overline{\zeta(t)}=0$ and $\overline{\zeta(t)\zeta(t')} = \delta(t,t')$), and the factor multiplying this variable results from the fluctuation-dissipation theorem. \textcolor{black}{ The parameters in the Langevin equation in Eq.\ (30) can be chosen to represent specific systems, such as in the Marcus-Levich electron transfer model\cite{chen2017electron}, single-atom heat engines\cite{ochoa2018quantum}, and molecular junctions with electronic friction \cite{dou2018perspective}. In the following examples, we will assume that the dominant relaxation mechanism for the primary phonons originates on their interaction with an equilibrium thermal bath, and will omit the potetial back-action from the electron dynamics.}

In the stationary state,(i.e., $t\to \infty$)
\begin{align}
  \overline{x_\alpha(t)} \propto{}& A_o \cos(\omega_d t - \phi),\\
  \overline{\left(x_\alpha (t) - \overline{x_\alpha(t)}\right)^2 }  \propto{} & \frac{k_B T_\alpha }{m_\alpha \omega_\alpha} .
\end{align}

where the phase factor $\phi$ depends on $\omega_\alpha$, $\omega_d$, and $\gamma$. The standard protocol to quantize the harmonic oscillator defines the mapping  $x_\alpha * \sqrt{2 m_\alpha \omega_\alpha / \hbar} \to  \hat X_\alpha$. These considerations allow us to assume the generic forms
\begin{align}
  \langle X_\alpha(t) \rangle =& c_1 \cos(\omega_d t + \phi_1),\\
  \langle X_\alpha^2(t) \rangle =&  c_2,
\end{align}
with $c1$, $c2$, and $\phi_1$ model constants, and restrict our study to situations with zero phonon cross correlation.

Figures \ref{fig:dot} and \ref{fig:dotM} show the sequential charging and discharging of a single dot subject to periodic driving on the level energy
\begin{equation}
  \label{eq:enet}
  \varepsilon (t) =   \varepsilon_o \cos(\omega_d t ),
\end{equation}
\textcolor{black}{where we have assumed that the energy level driving frequency is the same as the one for the periodic modulation in the phonon, Eq.\ \eqref{eq:mdtx}. In this form, the phonon and electron modulation rates are the same.}
We illustrate the dot model in Fig.\ \ref{fig:dot}(a).  The dot couples symmetrically to left and right reservoirs, $\Gamma_L = \Gamma_R= 0.01$ eV,  and is driven across a bias window defined by a difference in the chemical potential between the left and right reservoirs ($\mu_L= 0.05$ eV and $\mu_R = -0.05$ eV). The reservoirs' temperature is set to 10 K, and the dot level energy is initially set to $-0.1$ eV, having the dot initially charged ($\rho(0) = 1.0$). First, we consider a phonon-free system (i.e. $M=0$) and a process with a driving frequency $\omega_d = 2$ meV in Figs.\ \ref{fig:dot}(b) and (c). We report the transient left and right currents in Fig.\ \ref{fig:dot}(b), and show the variation in level occupation and energy in Fig.\ \ref{fig:dot}(c). The current is positive when charge is transferred from the reservoir to the dot. Relative to the driving frequency, we observe a quick but short drop in the initial electron occupation in response to the dot coupling to the reservoirs. This results in a small current from the system to the right reservoir, and a slight drop in the electron population. After this, the electron occupation and the currents vary periodically, going from a nearly fully occupied to a nearly depleted dot as the energy level changes from $-0.1$ to $0.1$ eV. The left and right currents both increase in magnitude when the level energy falls within the bias window,  $\mu_L > \varepsilon(t) > \mu_R$, and drop otherwise. When the level energy is in resonance with one reservoir's Fermi energy, the current from such reservoirs displays fast oscillations, resulting from the competing time scales associated with the level driving and the dot-reservoir electron exchange.

Next, we investigate the response of the dot when it couples to a phonon in Fig.\ \ref{fig:dotM}. In this example, we assume that the phonon is periodically driven at the same frequency than the level energy (i.e., $\omega_d = 2 meV$), such that the phonon and electron dynamics are in the same timescale. In Fig.\ \ref{fig:dotM}(a) we observe the variation in the transient currents as we increase the electron-phonon coupling strength $M$ from 0 to 0.035 eV.  When $M < \Gamma_{L/R}$, we find that the current profile is not sensitive to the presence of the driven phonon and does not deviate significantly from the one observed on the free dot. On the contrary, the phonon dynamics modify $I_L$ and $I_R$ when $M>\Gamma_{L/R}$. Specifically, in the case $M=0.015$ eV, we note a small suppression of the fast oscillations observed in Fig.\ \ref{fig:dot}(b) when the level is in resonance with one reservoir's Fermi energy (cf. gray and brown lines in Fig. \ref{fig:dotM}(a)). When $M = 0.025$ eV, the classical phonon broadens the level energy allowing for increased currents, relative to the free dot,  when the level is outside the bias window. Moreover, the fast oscillations near resonance are completely suppressed. The current profile when $M=0.035$ eV shows that $I_L$ and $I_R$ are nearly time independent when the electron-phonon coupling is strong relative to the coupling to the baths. In this case, the current magnitude is close to the maximum observed on the free dot. Figure \ref{fig:dotM}(b) shows the transient electron population in the dot for the same set of $M$ values, finding that as $M$ increases the dot cyclic transition from depleted to fully occupied breaks, achieving a nearly time-independent half-occupied dot when $M \gg \Gamma_{L/R}$.

While our model describes the phonon classically, we rationalize the results in Figs.\ \ref{fig:dot} and \ref{fig:dotM} in terms of the emergence of hybrid electron-phonon states as $M$ increases. Indeed, from a fully quantum picture of the system, electron-phonon couplings lead to the emergence of multiple channels for electron transport. In the limit of a large number of active phonon modes, the phonon is well represented by the classical picture whenever the discrete transport state is smeared forming a single broad level. We conjecture that the smearing extension is indeed proportional to the coupling strength and the phonon driving. 

\begin{figure}[th]
  \centering
  \includegraphics[scale=0.49]{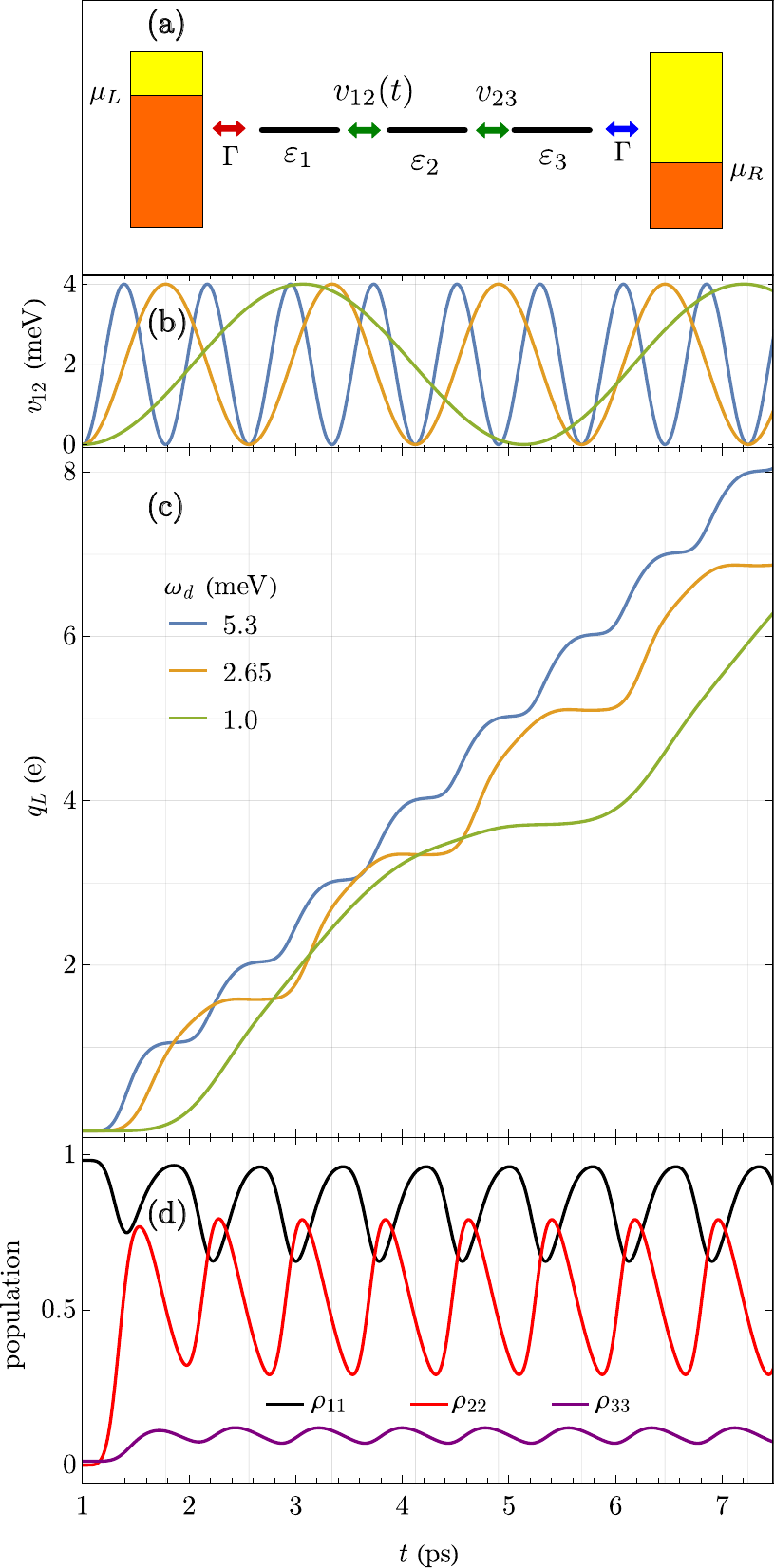}
  \caption{A model for a single-electron pump. (a) Sketch of the system which consisting of a three sequentially-coupled energy levels. Counting from the right to the left, the tunneling energy between the first and second levels $v_{12}$ is modulated periodically from zero to $2 v_{23}$. The first level is coupled to an electron reservoir with chemical potential $\mu_L$ and the third level is coupled to the right reservoir with chemical potential $\mu_R$. The systems is defined so that $\mu_L > \varepsilon_1 = \varepsilon_2 = \varepsilon_3 > \mu_R$, and $T_R = T_L$. (b) $v_{12}$ variation for various modulation frequencies $\omega_d$: $1.0$ meV (green), $2.65$ meV (yellow), and  $5.3$ meV (blue). (c) Net charged transferred from the left reservoir to the system as function of time and for the three different driving schemes in (b). (d) electron population variation as a function as a function of time for the single electron pump protocol, $\omega_d =5.3$ meV. Other parameters for this system are $\mu_L=0.1$ eV, $\mu_R= -0.1$ eV, $\varepsilon_1=\varepsilon_2=\varepsilon_3=0$ eV, $v_{23}=2 $ meV, $\Gamma =0.01$ eV, $T_R= T_L = 10$ K.    }
  \label{fig:1epump}
\end{figure}

\begin{figure}[th]
  \centering
  \includegraphics[scale=0.68]{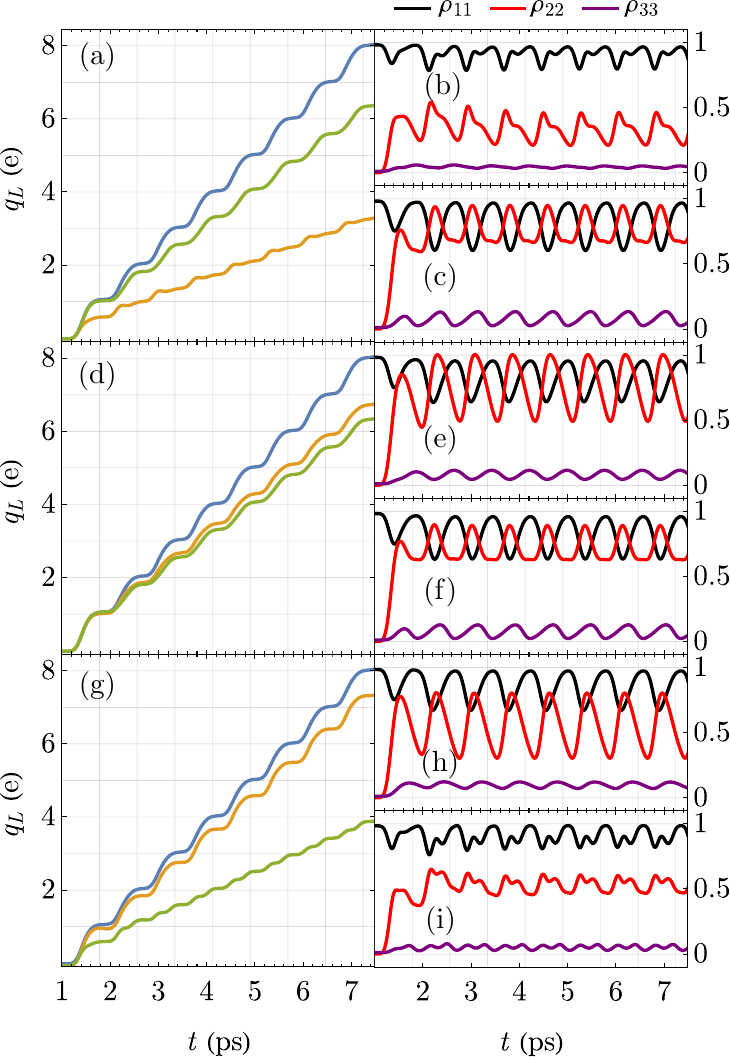}
  \caption{Single-electron pumps with electron phonon-couplings. (a), (d) and (g) show the total charge transferred from the left reservoir into the system for the protocol described in Fig.\ \ref{fig:1epump} with $\omega_d = 5.3$ meV and with different types of electron-phonon coupling. In each case, phonons are periodically driven with frequency $\omega_d$, but may differ on the phase factor $\phi_1$. Panels (b), (c), (e), (f), (h), and (i) display the electron level population $\rho_{11}$ (black), $\rho_{22}$ (red), and $\rho_{33}$ (purple) as follows. (a) a single phonon couples to level 2 with $\phi_1=\pi/2$ (orange, electron population in (b)) and $\phi_1=3 \pi/2$ (green, electron population in (c)). In (d) a single phonon couples to level 3 with $\phi_1=\pi/2$ (orange,, electron population in (e)) and $\phi_1=3 \pi/2$ (green,electron population in (f)). (g) A single phonon coupled to the three levels with $\phi_1=3\pi/2$  (orange,, electron population in (h)) and three independent phonons, each one coupled to a different level with relative phases $\phi_1({\rm level}$ $ 1)=\pi =\phi_1({\rm level}$ $ 3)$, $\phi_1({\rm level}$ $2)=0$ (green, electron population in (i)). Parameters $M=5$ meV, $c_1= 0.945$, $c_2= 35.71$. Other parameters are as in Fig.\ \ref{fig:1epump}. (a),(d) and (g) also include the phonon-free system (blue), already analyzed in \ref{fig:1epump}(c) for comparison.
  }
  \label{fig:1epumpM}
\end{figure}

Next, we investigate a model for a single electron pump in Figs.\ \ref{fig:1epump} and \ref{fig:1epumpM}. Single-electron sources are of interests for standards, such as the Ampere definition in terms of the elementary charge $e$ and transfer rate $f$, such that the net current is $I = e f$.  We introduce in Fig.\ \ref{fig:1epump}(a) a simple model for a single-electron pump consisting of a single-gate protocol with periodic modulation on the entrance gate potential\cite{pothier1992single,norimoto2024statistical}. The model consists of a three identical levels ($\varepsilon_1=\varepsilon_2=\varepsilon_3=0.0$), linearly coupled  with time-periodic modulation of the tunneling energy $v_{12}$ between levels 1 and 2 
\begin{align}\label{eq:v12}
  v_{12}(t) =& v_{23}\left[ 1 - \cos (\omega_d t)\right],
\end{align}
where $\omega_d$ is the gate driving frequency and $v_{23}$ is the tunneling between levels 2 and 3, which we fix during the protocol. Level 1 and 3 are correspondingly coupled  to fermionic left and right reservoirs, with $\mu_L > \mu_R$. By varying $\omega_d$, we find in Fig.\ \ref{fig:1epump} a protocol that effectively pumps a single electron per driving period from the left reservoir. Figure \ref{fig:1epump}(b) shows $v_{12}$ for three different values in the driving frequency $\omega_d$ ( $5.3, 2.65$, and $1.0$ meV) as a function of time. We calculate in Fig.\ \ref{fig:1epump}(c) the net charge pumped from the left reservoir by
\begin{equation}
  q_L(t) = \int_0^t I_L(t) dt \label{eq:qL},
\end{equation}
for the three different frequencies in Fig.\ \ref{fig:1epump}(b). In these calculations we assumed $\rho_{11}=1.0$ and $\rho_{22}=\rho_{33}=0.0$ at $t=0$, with $v_{12}(0) = 0$, and allowed the system to evolve for 1 ps before turning on the modulation in $v_{12}$ Eq.\ \eqref{eq:v12}.  The grid in Fig.\ \ref{fig:1epump}(c) marks with the vertical lines integer multiples of the period for driving frequency for $\omega_d = 5.3$ meV, and the horizontal lines correspond to integer multiples of the electron charge $e$. In this form, Fig.\ \ref{fig:1epump}(c) reveals that driving  periodically $\omega_d$ at the 5.3 meV (blue line) we effectively pump a single electron per driving period, illustrating also that driving at smaller frequencies results in more than one electron pumped during their corresponding period (green and yellow lines). As another interesting finding, we note that driving at half the optimal single-electron-pumping frequency (i.e. 2.65 meV) does not result in a process injecting two electrons during the corresponding period.

The classical representation of the pump\cite{yamahata2021understanding} envisions the process as divided in three stages: first the electron placed in level 1 remains there until $v_{12}$ is large enough to inject the electron from level 1 into level 2. During the second stage the electron waits in level 2 while $v_{12}$ drops blocking back injection, eventually favoring also the electron transport from level 2 into level 3. In the third stage the electron escapes from level 3 to the right reservoir and the process repeats. We investigate the validity of such description in our model for a single-electron pump in Fig.\ \ref{fig:1epump}(d), showing the electron population as function of time for each level. Our results indicates that for the system parameters and the driving protocol considered, such classical picture breaks. Instead, the expectation value to find an electron in each level   is always finite, displaying large oscillations in level 2, while level 1 and 3 are correspondingly nearly filled and nearly depleted during the pumping.

We can now investigate the effect of electron-phonon couplings in the performance of our single electron pump. In Fig.\ \ref{fig:1epumpM}, we consider several scenarios with phonons coupling in different forms to our three level system, while driven at the same frequency $\omega_d = 5.3$ meV. We also consider the effect of the phase $\phi_1$ in the phonon dynamics. Figure \ref{fig:1epumpM}(a) displays $q_L$, when a phonon couples to level 2 with two different phase factors tuning the phonon dynamics from in-phase ( $\phi_1 = \pi/2$ ) to anti-phase ($\phi_1 = 3 \pi/2$) with respect to the $v_{12}$. We observe that in both cases the electron-phonon coupling lowers the charge transferred from the left reservoir during the pumping process, having a more significant effect when the phonon is in resonance. We observe a major change in the population dynamics for the in-phase case (Fig.\ \ref{fig:1epumpM}(b)) relative to the phonon-free scenario, with a significant suppression of the oscillations in $\rho_{22}$ and $\rho_{11}$, suggesting that the net effect of the phonon dynamics is to partially block the electron injection from level 1 into level 2. In the anti-phase situation, we observe an enhanced electron density at level 2 and a drop on the mean electron population in level 3. Thus, we conclude that anti-phase phonon driving reduced the injection rate from level 2 to level 3.

Figures \ref{fig:1epumpM}(d), (e) and (f) show the impact of single phonon coupled to level 3. We observe again a lowering in $q_L$ per driving cycle. However, compared to the situation in which the phonon couples to level 2, the drop in the pumping efficiency is less significant. The in-phase and anti-phase protocols lead to very similar electron pumping rates, with the anti-phase case leading to slightly lower rate. Relative to the system without phonons, Figs.\ \ref{fig:1epumpM}(e) and (f) suggest that in both cases the phonon partially obstructs electron transfer from level 2 into level 3. Finally, we consider situations with multiple levels coupling to phonons in Figs.\ \ref{fig:1epumpM}(g), (h), and (i).Specifically, we analyze the situation with a single anti-phase driven phonon couples equally to each level, and when each level couples to an independent phonon with different phases.  The former case shows a minor change in $q_L$ relative to the phonon-free system. The latter, on the contrary, brings down the pumped charge to nearly half the one expected from the noninteracting system, showing that a highly disordered system in the phonon driving, is detrimental to the electron transfer rate.

We end this section with two remarks. First, we note that in our models for the quantum dot and the single-electron pump the timescale associated with the primary phonon dynamics was similar to the electron timescale. While the protocols considered here can be regarded as very fast for the current operational capabilities in actual devices, the models serve to illustrate the potential of the semiclassical Green's functions and their corresponding equations of motion in describing electron transport problems when the system's phonons dynamics is comparable with the electronic transitions. The second remark pertains on the back-action of the electron dynamics in the phonon, which we have not included in the Langevin equation. Such back-action can be anticipated from the phonon-Green-function equation of motion, or can be introduced by adopting other phenomenological models(see. Ref.\ \citenum{kershaw2020non} ). \textcolor{black}{Future work will look into this important aspect, as well as extending this work to include more complex phonon dynamics with colored noise and non-Markovianity.}

\section{Conclusion}\label{sec:conclusion}

We introduced a non-equilibrium Green's Function formalism to calculate transient electron transport in open nanoscale and molecular systems with electron-phonon couplings. Our method considers the expansion of the scattering matrix in powers of the electron-phonon coupling strengths, obtaining differential equations of motion for the lesser and retarded projections of the electronic Green's Functions. These equations are simplified by describing the system phonon dynamics by time-local classical or phenomenological stochastic equations with delta-correlated noise, and by adopting the wide-band approximation.  Numerical propagation of these equations allows us to compute electronic currents, electron density and electron coherence during the transient regime, incorporating the impact of the system's phonon dynamics on these observables. We illustrated the method by studying models for the charging and discharging of a quantum dot, and a single-electron pump. In both systems we assumed harmonic phonons driven periodically at the same frequency than the modulated electron degrees-of-freedom. \textcolor{black}{While the present approach is perturbative in the electron-phonon coupling strength, such as most quantum master equation methods, the semiclassical Green's function method provides a simple and systematic approach to introduce classical and phenomenological equations of motion for the phonon dynamics. Significantly, our method does not invoke assumptions such as the Born-Markov approximation, which limits the system's modulation timescale. Moreover, in terms of the size of the electronic system, the numerical propagation of the electronic Green's function is computationally more efficient.} Finally, we expect the present method to find applications in the  description of complex systems such as molecular junctions, molecular motors, and nanoelectromechanicals with fast electron and phonon dynamics.
\vspace{3cm}

\section*{Data Availability}

The data that support the findings of this study are available from the corresponding author upon reasonable request.

\bibliography{semiclass.bib}

\end{document}